\documentclass[aps,prd,preprint,preprintnumbers,amsmath,amssymb,superscriptaddress,showpacs,floatfix]{revtex4-1}

\usepackage{graphicx}
\usepackage{dcolumn}
\usepackage{bm}
\usepackage{color}
\usepackage{epsfig}
\usepackage{slashed}
\allowdisplaybreaks
\begin{document}

\title{Azimuthal asymmetries in semi-inclusive DIS with polarized beam and/or target and their nuclear dependences}
\author{Yu-kun Song}
 \affiliation{School of Physics, Shandong University, Jinan, Shandong 250100, China}
 \affiliation{Department of Modern Physics, University of Science and Technology of China, Hefei, Anhui 230026, China}
 \affiliation{Key Laboratory of Quark \& Lepton Physics (Central China Normal University), Ministry of Education, China}
\author{Jian-hua Gao}
 \affiliation{Shandong Provincial Key Laboratory of Optical Astronomy and Solar-Terrestrial Environment,  School of Space Science and Physics, Shandong University at Weihai, Weihai 264209, China}
 \affiliation{Key Laboratory of Quark \& Lepton Physics (Central China Normal University), Ministry of Education, China}
\author{Zuo-tang Liang}
 \affiliation{School of Physics, Shandong University, Jinan, Shandong 250100, China}
\author{Xin-Nian Wang}
 \affiliation{Key Laboratory of Quark \& Lepton Physics (Central China Normal University), Ministry of Education, China}
 \affiliation{Institute of Particle Physics, Central China Normal University, Wuhan, 430079, China}
 \affiliation{Nuclear Science Division, MS 70R0319, Lawrence Berkeley National Laboratory, Berkeley, California 94720}

\date{\today}


\begin{abstract}
Using the formalism obtained from collinear expansion, 
we calculate the differential cross section and azimuthal asymmetries in semi-inclusive 
deeply inelastic lepton-nucleon (nucleus) scattering process $e^-+N (A) \to e^-+q+X$ 
with both polarized beam and polarized target up to twist-3. 
We derive the azimuthal asymmetries in terms of twist-3 parton correlation functions. 
We simplify the results by using the QCD equation of motion that leads to a set of 
relationships between different twist-3 functions. 
We further study the nuclear dependence of azimuthal asymmetries and 
show that they have similar suppression factors as those in the unpolarized reactions. 

\end{abstract}


\maketitle


\section{Introduction}
Azimuthal asymmetries in semi-inclusive deeply inelastic lepton-nucleon scattering (SIDIS) play 
an important role in the study of partonic structure of nucleon, attracting much effort in both theory~\cite{Georgi:1977tv,Cahn:1978se,Berger:1979kz,Liang:1993re,Mulders:1995dh,Oganesian:1997jq,Chay:1997qy,Nadolsky:2000kz,Bacchetta:2006tn,Anselmino:2006rv,Anselmino:2005nn,Liang:2006wp,Gao:2010mj,Song:2010pf,Gao:2011mf} and experiment~\cite{Aubert:1983cz,Arneodo:1986cf,Adams:1993hs,Breitweg:2000qh,Chekanov:2002sz,Airapetian:2001eg,Airapetian:2002mf,Airapetian:2004tw,Alexakhin:2005iw,Webb:2005cd,Osipenko:2008aa,Alekseev:2010dm,Airapetian:2012yg}.
In such studies, spin and nuclear dependences are often important and provide 
an useful tool to investigate these effects.  
Also because of this, higher twist contributions are often significant and need to be taken into account precisely.
Besides, such higher twist effects usually depend on new higher twist parton correlation functions 
hence the studies of them provide a new window to learn about the structure of nucleon. 

One of the most important issues in these studies is to establish the relationships between the experimentally
measurable quantities and different parton distribution and/or correlation functions that describe
the partonic structure of the nucleon and the properties of the hadronic interaction in a consistent theoretical framework.
Collinear expansion seems to be the most promising technique that leads to such a framework.
It was proposed in 1980s and has been successfully applied to 
the inclusive processes~\cite{Ellis:1982wd,Ellis:1982cd,Qiu:1988dn,Qiu:1990xxa,Qiu:1990xy}.
It has been shown that, after collinear expansion, the differential cross section can be expressed as
a convolution of the collinear expanded hard parts with the parton distribution 
and/or correlation functions in nucleon.
While the hard parts are calculable, 
the parton distribution and/or correlation functions can be defined in terms of 
gauge invariant matrix elements of the nucleon state.
These matrix elements contain the information about parton distributions inside the nucleon. The gauge link inside the gauge invariant matrix 
elements is a result of multiple gluon scattering within the collinear expansion scheme. 
Within this scheme, one performs Taylor expansion of the hard parts around the collinear momenta. 
The leading twist contributions come from the zeroth order in the collinear expansion allowing 
all momenta taking their collinear values.  
Higher twist contributions from the higher orders of the Taylor expansion can be calculated  consistently.

Higher twist effects in semi-inclusive deeply inelastic lepton-nucleon scattering have also been studied 
in literature and calculations of the differential cross section 
up to twist-3 level have been carried out~\cite{Mulders:1995dh,Oganesian:1997jq,Nadolsky:2000kz,Bacchetta:2006tn}. 
However, most of these studies do not consider the application of collinear expansion. 
In stead, they usually start from the expressions obtained directly from the Feynman diagrams, 
extract the leading (twist-2) and the sub-leading (twist-3) twist contributions by making 
appropriate approximations, and insert the gauge link whenever needed to guarantee the gauge 
invariance of the parton distribution and/or correlation functions. 
It is thus unknown whether, if yes, how the collinear expansion is applicable here. 
It is not obvious where the gauge link comes from and which form it takes. 
It is also not known whether the calculations extend to even higher twist. 
A systematic study leading to a consistent formalism is necessary but still lacking. 

In Ref.~\cite{Liang:2006wp}, we made the first step towards this goal by applying 
the collinear expansion technique to the semi-inclusive deep inelastic scattering (SIDIS)
process $e^-+N\to e^-+q+X$,
where $q$ denotes a quark which is equivalent to a jet in experiment.
We showed that the collinear expansion technique is applicable for this process and 
derived a formalism suitable for studying leading as well as higher twist contributions 
to $e^-+N\to e^-+q+X$ in a systematic way.  
This formalism is similar to that we have for inclusive process 
and similar expressions can be obtained for the differential cross section 
or the hadronic tensor as a convolution of the hard parts and 
the un-integrated or transverse momentum dependent (TMD)
parton correlation functions.
We carried out the calculations of the azimuthal asymmetries in the unpolarized cases 
up to twist-4~\cite{Song:2010pf} and those in the case with transversely polarized targets 
up to twist-3~\cite{Liang:2006wp}.
Furthermore, we also showed that the multiple gluon scattering contained in the gauge link 
leads to a significant nuclear dependence of the azimuthal asymmetries which can be studied experimentally~\cite{Liang:2008vz,Gao:2010mj}.

In this paper,  we present calculations of azimuthal asymmetries 
in the semi-inclusive process $e^-+N (A) \to e^-+q+X$
with beam and target in different polarizations up to twist-3  
using the formalism derived in \cite{Liang:2006wp}.
For completeness, we summarize the formalism in Sec. II and present the results of the hadronic tensor.
In Sec. III, we present the results of the differential cross sections and the azimuthal asymmetries.
We study the nuclear dependence in Sec. IV and conclude in Sec. V.

\section{The hadronic tensor}
The formalism that we use in our calculations are derived in~\cite{Liang:2006wp} for the semi-inclusive
process $e^-+N\to e^-+q+X$ and are summarized in~\cite{Song:2010pf}.
It is obtained by applying collinear expansion and contains the contributions from the multiple gluon scattering.
For completeness and also for comparison with other approaches such as those in~\cite{Mulders:1995dh,Bacchetta:2006tn}, 
we summarize the most related results of this formalism in part A and
present the results for the hadronic tensors in different polarized cases up to twist-3 
in other parts of this section.

\subsection{The formalism}

We consider the SIDIS process $e^-+N\to e^-+q+X$ and use 
$l$, $l'$, $p$, $k$ and $k'$ to denote the four-momenta of electron, 
nucleon and parton respectively, those with primes are for the final state.
The polarization vectors are denoted by $s_l$ and $s$ and are taken as 
$s_l^\mu=\lambda_l l^\mu/m_e+s_{l\perp}^\mu$, and $s^\mu=\lambda p^\mu/M+s_{\perp}^\mu$, 
where $\lambda_l$ and $\lambda$ are the helicities.
We use light-cone coordinate $k^\mu=(k^+,k^-,\vec k_\perp)$ and take
unit vectors as $\bar n=(1,0,\vec 0_\perp)$, $n=(0,1,\vec 0_\perp)$, and $n_{\perp}=(0,0,\vec n_\perp)$.
We choose the coordinate system such that,
$p=p^+\bar n$, $q=-x_Bp+nQ^2/(2x_Bp^+)$, $l_\perp=|\vec l_\perp|n_{\perp 1}$, 
where $x_B=Q^2/2p\cdot q$ and define $y=p\cdot q/p\cdot l$.

The differential cross section is given by,
\begin{equation}
d\sigma=\frac{2\alpha_{\rm em}^2e_q^2}{sQ^4}L^{\mu\nu}(l,l',s_l)\frac{d^2W_{\mu\nu}}{d^2k'_\perp}
\frac{d^3l'd^2k'_\perp}{2E_{l'}},
\end{equation}
where $L^{\mu\nu}(l,l^\prime,s_l)$ is the leptonic tensor,  and 
the hadronic tensor is given by,
\begin{align}
  &\frac{d^2W_{\mu\nu}}{d^2k_\perp^\prime}=\int\frac{dk_z^\prime}{(2\pi)^32E_{k^\prime}}W^{({\rm si})}_{\mu\nu}(q,p,s,k^\prime),\\
  &W_{\mu\nu}^{({\rm si})}(q,p,s,k^\prime)=\frac{1}{2\pi}\sum_X\langle p,s\left|j_\mu(0)\right|k^\prime,X\rangle\langle k^\prime,X\left|j_\nu(0)\right|p,s\rangle(2\pi)^4\delta^4(p+q-p_X).
\end{align}

As discussed in \cite{Ellis:1982wd,Ellis:1982cd} for inclusive DIS and in \cite{Liang:2006wp} 
for semi-inclusive DIS, to obtain the gauge invariant form of the hadronic tensor including 
leading and higher twist contributions in a systematic way, 
at the tree level, we need to consider the Feynman diagram series as illustrated in Fig.1.  
The hadronic tensor is given by a sum of the contribution from each diagram $W_{\mu\nu}^{({\rm si})}=\sum_jW_{\mu\nu}^{(j,{\rm si})}$.
For example, for $j=0,1$ and 2, 
\begin{align}
&W_{\mu\nu}^{(0,{\rm si})}=\frac{1}{2\pi}
\int\frac{d^4k}{(2\pi)^4} 
{\rm Tr}[\hat H_{\mu\nu}^{(0)}(k,q)\hat \phi^{(0)}(k,p,S)] 2E_{k'}(2\pi)^3\delta^3(\vec k'-\vec k-\vec q),\label{eq:Wsi0}\\
&W_{\mu\nu}^{(1,{\rm si})}= \frac{1}{2\pi}\int\frac{d^4k_1}{(2\pi)^4}\frac{d^4k_2}{(2\pi)^4}
\sum_{c=L,R} 2E_{k'}(2\pi)^3\delta^3(\vec k'-\vec k_c-\vec q)\nonumber\\
&\phantom{XXXXXXXXXXXXX}\times {\rm Tr}[\hat H_{\mu\nu}^{(1,c)\rho}(k_1,k_2,q) 
\hat \phi^{(1)}_\rho(k_1,k_2,p)], \label{eq:Wsi1}\\
&W_{\mu\nu}^{(2,{\rm si})}= \frac{1}{2\pi}\int\frac{d^4k_1}{(2\pi)^4}\frac{d^4k_2}{(2\pi)^4}\frac{d^4k}{(2\pi)^4}
\sum_{c=L,R,M} 2E_{k'}(2\pi)^3\delta^3(\vec k'-\vec k_c-\vec q)\nonumber\\
&\phantom{XXXXXXXXXXXX}\times {\rm Tr}[\hat H_{\mu\nu}^{(2,c)\rho\sigma}(k_1,k_2,k,q) 
\hat \phi^{(2)}_{\rho\sigma}(k_1,k_2,k,p,S),  \label{eq:Wsi2}
\end{align}
where $c$ denotes different cuts, $k_L=k_1, k_R=k_2$, $k_M=k$,
and the hard parts are given by,
\begin{align}
&\hat H_{\mu\nu}^{(0)}(q,k)=
\gamma_\mu(\slash{\hspace{-5pt}k}+\slash{\hspace{-5pt}q})\gamma_\nu
(2\pi)\delta_+((k-q)^2),  \label{eq:h0}\\
&\hat H_{\mu\nu}^{(1,L)\rho}(k_1,k_2,q)=
\gamma_\mu (\slash{\hspace{-5pt}k_2}+\slash{\hspace{-5pt}q})\gamma^\rho
\frac{\slash{\hspace{-5pt}k_1}+\slash{\hspace{-5pt}q}}
{(k_1+q)^2-i\epsilon} \gamma_\nu (2\pi)\delta_+((k_2+q)^2),  \label{eq:h1}\\
&\hat H_{\mu\nu}^{(2,L)\rho\sigma}(k_1,k_2,k,q)=
\gamma_\mu (\slash{\hspace{-5pt}k_2}+\slash{\hspace{-5pt}q})\gamma^\rho
\frac{\slash{\hspace{-5pt}k}+\slash{\hspace{-5pt}q}}
{(k+q)^2-i\epsilon}\gamma^\sigma \frac{\slash{\hspace{-5pt}k_1}+\slash{\hspace{-5pt}q}}
{(k_1+q)^2-i\epsilon} \gamma_\nu (2\pi)\delta_+((k_2+q)^2),  \label{eq:h2}
\end{align}
and the matrix elements or the correlators are defined as,
\begin{align}
& \hat\phi^{(0)}(k,p,S) =  
\int d^4ze^{ikz} \langle p,S|\bar\psi(0)\psi(z)|p,S\rangle, \label{eq:phi0}\\
&\hat\phi^{(1)}_\rho(k_1,k_2,p,S) =  
\int d^4yd^4ze^{ik_1z+i(k_2-k_1)y} \langle p,S|\bar\psi(0)gA_\rho(z)\psi(y)|p,S\rangle, \label{eq:phi1}\\
& \hat\phi^{(2)}_{\rho\sigma}(k_1,k_2,k,p,S) =
\int d^4yd^4y'd^4z e^{ik_1\cdot y+i(k-k_1)\cdot z^\prime+i(k_2-k)\cdot z}   \nonumber\\
&\phantom{XXXXXXXXXXXXXXXX}\times \langle p,S|\bar\psi(0) gA_\rho(z)gA_\sigma(z^\prime) \psi(y)|p,S\rangle. \label{eq:phi2}
\end{align}

\begin{figure}[h!]
 \centering
  \epsfig{file=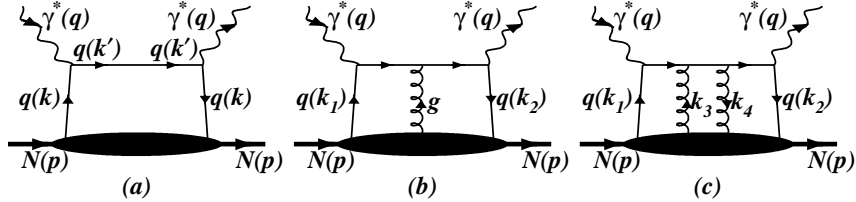,width=0.8\textwidth,clip=}
  \caption{Examples of the Feynman diagram series considered 
  for $\gamma^*+N\to q+X$ with (a) $j=0$, (b) $j=1$ and (c) $j=2$ gluons exchanged.}
\end{figure}

As it is well known that, since the field operators in the correlators given by Eqs. (\ref{eq:phi0}-\ref{eq:phi2})
are at different space time points, these correlators and the parton distribution and/or correlation 
functions defined from them are not gauge invariant. 
To reach the gauge invariant form, we need to do the collinear expansion. 
The expansion has been carried out and summarized in \cite{Liang:2006wp,Song:2010pf}. 
We present the main results in the following.   
For brevity, we show only $j=0$ and $1$ terms 
that are needed in our calculations in this paper up to twist-3.

It has been shown that~\cite{Liang:2006wp}, after collinear expansion, the hadronic tensor takes the form,
\begin{align}
  \frac{d^2W_{\mu\nu}}{d^2k_\perp^\prime}=&
  \frac{d^2\tilde W_{\mu\nu}^{(0)}}{d^2k_\perp^\prime}+\frac{d^2\tilde W_{\mu\nu}^{(1)}}{d^2k_\perp^\prime}+\dots,\label{dwdk:collinear-expanded} \\
\frac{d\tilde W_{\mu\nu}^{(0)}}{d^2k'_\perp}=&\frac{1}{2\pi} \int dx d^2k_\perp
{\rm Tr}\big[\hat H_{\mu\nu}^{(0)}(x)\ \hat \Phi^{(0)N}(x,k_\perp)\big] \delta^{(2)}(\vec k_\perp-\vec{k'}_\perp),\label{eq:tWsi0}\\
\frac{d\tilde W_{\mu\nu}^{(1)}}{d^2k'_\perp}=&\frac{1}{2\pi}  \int dx_1d^2k_{1\perp} dx_2d^2k_{2\perp} \nonumber\\
&\times \sum_{c=L,R}
{\rm Tr}\bigl[\hat H_{\mu\nu}^{(1,c)\rho}(x_1,x_2) \omega_\rho^{\ \rho'}
\hat \Phi^{(1)N}_{\rho'}(x_1,k_{1\perp},x_2,k_{2\perp})\bigr] \delta^{(2)}(\vec k_{c\perp}-\vec{k'}_\perp), \label{eq:tWsi1}
\end{align}
where the tilded symbols $\tilde W^{(j)}$'s represent results after collinear expansion and 
$\omega_\rho^{\ \rho'}=g_\rho^{\ \rho'}-\bar n_\rho n^{\rho'}$ is a projection operator.
The matrix elements take the gauge invariant form and are given by,
\begin{align}
&\hat\Phi^{(0)N}(x,k_\perp)=\int \frac{p^+dy^- d^2y_\perp}{(2\pi)^3}  e^{ix p^+ y^- -i\vec{k}_{\perp}\cdot \vec{y}_{\perp}}
\langle N|\bar{\psi}(0){\cal{L}}(0;y)\psi(y)|N\rangle, 
\label{eq:Phi0def}\\
&\hat\Phi^{(1)N}_\rho(x_1,k_{1\perp},x_2,k_{2\perp})
=\int \frac{p^+dy^- d^2y_\perp}{(2\pi)^3}\frac{p^+dz^-d^2z_\perp}{(2\pi)^3} \nonumber\\ 
&~~~~~~\times e^{ix_2p^+z^- -i\vec{k}_{2\perp}\cdot \vec{z}_{\perp}+ix_1p^+(y^--z^-)-i\vec{k}_{1\perp}\cdot (\vec y_\perp-\vec z_\perp)} 
\langle N|\bar\psi(0) {\cal L}(0;z)D_\rho(z){\cal L}(z;y)\psi(y)|N\rangle,
\label{eq:Phi1def}
\end{align}
where ${\cal{L}}(0;y)$ is the gauge link obtained in the collinear expansion and is given by, 
\begin{eqnarray}
\label{CollinearGL}
&&{\cal{L}}(0;y)=\mathcal{L}^\dag_{\parallel}(\infty,\vec{0}_\perp; 0,\vec{0}_\perp)
\mathcal{L}_\perp(\infty;\vec 0_\perp,\vec{y}_{\perp})
\mathcal{L}_{\parallel}(\infty,\vec{y}_{\perp}; y^-,\vec{y}_{\perp}),\\
\label{TMDGL}
&&{\cal{L}}_{\parallel}(\infty,\vec{y}_{\perp}; y^-,\vec{y}_{\perp})=
P e^{- i g \int_{y^-}^\infty d \xi^{-} A^+ ( \xi^-, \vec{y}_{\perp})} , \\
&& \mathcal{L}_\perp(\infty;\vec 0_\perp,\vec{y}_{\perp})=
P e^{-ig\int_{\vec 0_\perp}^{\vec y_\perp} d\vec\xi_\perp\cdot\vec A_\perp(\infty,\vec\xi_\perp)},
\end{eqnarray}
where $P$ stands for path integral. 
The hard parts in these tilded $\tilde W^{(j)}$'s are the first terms in the Taylor expansions 
at $k_i=x_ip$ of the corresponding hard parts obtained directly from the Feynman diagrams. 
They are given by~\cite{Liang:2006wp},
\begin{align}
&\hat H_{\mu\nu}^{(0)}(x)=\frac{2\pi}{2q\cdot p}
\gamma_\mu\big(\slash{\hspace{-5pt}q}+x\slash{\hspace{-5pt}p}\big)\gamma_\nu\delta(x-x_B),\\
&\hat H_{\mu\nu}^{(1,L)\rho}(x_1,x_2)=\frac{2\pi}{(2q\cdot p)^2}
\frac{\gamma_\mu\big(\slash{\hspace{-5pt}q}+x_2\slash{\hspace{-5pt}p}\big)\gamma^\rho\big(\slash{\hspace{-5pt}q}+x_1\slash{\hspace{-5pt}p}\big)
\gamma_\nu}{x_2-x_B-i\varepsilon}\delta(x_1-x_B).\label{hardpart:collinear-expanded}
\end{align}

These equations (\ref{dwdk:collinear-expanded}-\ref{hardpart:collinear-expanded}) 
form the basis for calculating the hadronic tensor in $e^-+N\to e^-+q+X$ at the tree level 
including leading and higher twist contributions. 
We emphasize once more that these equations are derived from the Feynman diagram series 
in Fig.1 using collinear expansion.   
They are nothing else but a reorganization of $W^{(j,{\rm si})}$ given by 
Eqs.~(\ref{eq:Wsi0}-\ref{eq:Wsi2}) obtained directly from this diagram series.  
We also note that $\tilde W^{(j)}$ differs distinctly from the corresponding $W^{(j,{\rm si})}$ 
and shows in particular the following features. 

(1) None of the tilded $\tilde W^{(j)}$ corresponds to one single Feynman diagram 
in the diagram series given in Fig.1. 
It contains contributions from all the infinite number of diagrams in this diagram series 
with exchange of $j=0$, 1, 2, ... gluon(s).

(2) The correlators acquire automatically the gauge links and are gauge invariant.  
The gauge link comes from the multiple gluon scattering shown in Fig.1. 
Furthermore, in the quark-gluon-quark correlator, covariant derivative is obtained 
to replace the gluon field operator in the original correlator before collinear expansion.  

(3) All the parton momenta in the hard parts take only the $\bar n$-components, 
while the corresponding $n$ and $n_\perp$ components are taken as zero.
Also there are projection operators $\omega_\rho^{\ \rho'}$'s 
in the expressions for $\tilde W^{(j)}$ for $j>0$ due to the collinear expansion. 

Because of the features mentioned above in particular point (3), the expressions for 
$\tilde W^{(j)}$ can be further simplified to a great deal. 
In fact, because of (3), the hard parts reduce to the following simple form, 
\begin{eqnarray}
&&\hat H^{(0)}_{\mu\nu}(x)=\pi\hat h^{(0)}_{\mu\nu}\delta(x-x_B),
\label{eq:H0simple}\\
&&\hat H^{(1,L)\rho}_{\mu\nu}(x_1,x_2)\omega_\rho^{\ \rho'}
=\frac{\pi}{2q\cdot p}\hat h^{(1)\rho}_{\mu\nu}\omega_\rho^{\ \rho'}\delta(x_1-x_B),
\label{eq:H1Lsimple}
\end{eqnarray}
where $\hat h^{(0)}_{\mu\nu}=\gamma_\mu\slash{\hspace{-5pt}n}\gamma_\nu/p^+$, and 
$\hat h^{(1)\rho}_{\mu\nu}=\gamma_\mu\slash{\hspace{-5pt}\bar n}\gamma^\rho\slash{\hspace{-5pt}n}\gamma_\nu$. 
We see not only that their $x$-dependences are contained only in the $\delta$-functions 
but also that $\hat H^{(1,L)\rho}_{\mu\nu}(x_1,x_2)\omega_\rho^{\ \rho'}$ depends only 
on $x_1$ but not on $x_2$.  
This implies that we can simply carry out the integration over $k_2$ in the correlator $\hat\Phi^{(1)}$ 
in $\tilde W^{(1)}$ and obtain, 
\begin{eqnarray}
\frac{d^2\tilde W^{(0)}_{\mu\nu}}{d^2k_\perp} &=&
\frac{1}{2}{\rm Tr}\bigl[\hat h^{(0)}_{\mu\nu}\hat\Phi^{(0)N}(x_B,k_\perp)\bigr],
\label{eq:W0} \\
\frac{d^2\tilde W^{(1,L)}_{\mu\nu}}{d^2k_\perp} &=&
\frac{1}{4q\cdot p}{\rm Tr}\bigl[\hat h^{(1)\rho}_{\mu\nu}\omega_\rho^{\ \rho'}\hat\varphi^{(1)N}_{\rho'}(x_B,k_\perp)\bigr],
\label{eq:W1L} 
\end{eqnarray}
The new correlator $\hat\varphi^{(1)}_\rho$ is defined as, 
$\hat\varphi^{(1)N}_\rho(x_1,k_{1\perp})\equiv \int dx_2d^2k_{2\perp}\hat\Phi^{(1)N}_\rho(x_1,k_{1\perp},x_2,k_{2\perp})$, and is given by,
\begin{eqnarray}
&&\hat\varphi^{(1)N}_\rho(x,k_\perp) =\int \frac{p^+dy^- d^2y_\perp}{(2\pi)^3}e^{ixp^+y^- -i\vec{k}_{\perp}\cdot \vec{y}_{\perp}}\langle N|\bar\psi(0) D_\rho(0){\cal L}(0;y)\psi(y)|N\rangle. \label{eq:varphi1}
\end{eqnarray}
It depends only on one parton momentum $k$, the quark field operator $\bar\psi$ and 
the covariant derivative $D_\rho$ are at the same space-time point.
Here, we may note that, unlike what we do in the current approach,  
the calculations presented in e.g. \cite{Bacchetta:2006tn} start from the hadronic 
tensors $W^{(j)}$'s given by Eqs. (\ref{eq:Wsi0}) and (\ref{eq:Wsi1}) obtained 
directly from the Feynman diagrams given by Figs.~1(a) and (b). 
To obtain the corresponding results, they need to make approximations for the hard parts by keeping 
only $\bar n$-terms, insert the gauge links into the matrix elements to guarantee the gauge invariance. 
Such operations are avoided in the formalism obtained using collinear expansion 
where the Feynman diagram series is considered systematically and the gauge links are obtained automatically.

\subsection{Twist-3 contributions to ${d^2W_{\mu\nu}}/{d^2k_\perp}$}

Up to twist-3, we need to consider the contributions from ${d^2\tilde W_{\mu\nu}^{(0)}}/{d^2k_\perp}$ and those from ${d^2\tilde W_{\mu\nu}^{(1)}}/{d^2k_\perp}$. 
Since the hard part $\hat h^{(0)}_{\mu\nu}$ and $\hat h^{(1)\rho}_{\mu\nu}$ have odd number of $\gamma-$matrices, 
only $\gamma^\alpha$ and $\gamma^\alpha\gamma_5$ terms of correlation matrices contribute. 
We decompose the correlation matrices as
  $\hat\Phi^{(0)}=\big(\Phi^{(0)}_\alpha\gamma^\alpha-\tilde \Phi^{(0)}_\alpha\gamma_5\gamma^\alpha\big)/2+ ...$,
  $\hat\varphi^{(1)}_{\rho}=\big(\varphi^{(1)}_{\rho\alpha}\gamma^\alpha-\tilde\varphi^{(1)}_{\rho\alpha}\gamma_5\gamma^\alpha\big)/2+ ...$,
and obtain the hadronic tensors as, 
\begin{align}
  \frac{d^2\tilde W^{(0)}_{\mu\nu}}{d^2k_\perp}&=\frac{1}{4}{\rm Tr}\big[\hat h^{(0)}_{\mu\nu}\gamma^\alpha\big]\Phi^{(0)}_\alpha-\frac{1}{4}{\rm Tr}\big[\hat h^{(0)}_{\mu\nu}\gamma_5\gamma^\alpha\big]\tilde\Phi^{(0)}_\alpha,\label{dwdk-0}\\
  \frac{d^2\tilde W^{(1,L)}_{\mu\nu}}{d^2k_\perp}&=\frac{1}{8p\cdot q}{\rm Tr}\big[\hat h^{(1)\rho}_{\mu\nu}\gamma^\alpha\big]\varphi^{(1)}_{\rho\alpha}-\frac{1}{8p\cdot q}{\rm Tr}\big[\hat h^{(1)\rho}_{\mu\nu}\gamma_5\gamma^\alpha\big]\tilde\varphi^{(1)}_{\rho\alpha}.\label{dwdk-1}
\end{align}

To proceed, we need to decompose the matrix elements involved in terms of the 
Lorentz covariants constructed from $p$, $n$, $k_\perp$ and $S$ multiplied by scalar functions 
of  $x$ and $k_\perp^2$.  
These scalar functions are just different components of the parton distribution and/or correlation functions. 
Such decompositions are the same as those discussed in different publications~\cite{Goeke:2005hb,Bacchetta:2006tn}. 
By inserting them into the above mentioned Eqs. (\ref{dwdk-0}) and (\ref{dwdk-1}), we can obtain the hadronic tensors 
in terms of these parton distribution and correlation functions. 
In the following, we calculate different contributions term by term. 
We first consider ${d^2\tilde W^{(0)}_{\mu\nu}}/{d^2k_\perp}$. 
Up to twist-3 level, $\Phi^{(0)}_\alpha$ and $\tilde\Phi^{(0)}_\alpha$ 
are decomposed as~\cite{Goeke:2005hb},
\begin{align}
\displaystyle\Phi^{(0)}_{\alpha}=&
\left(f_1-\varepsilon_\perp^{ks}f_{1T}^\perp\right)p_\alpha+f^\perp k_{\perp \alpha}
-f_TM\varepsilon_{\perp\alpha i}s_\perp^i\nonumber\\
&-\frac{f_T^\perp}{M}\Big(k_{\perp\alpha}k_{\perp\beta}-\frac{1}{2}k_\perp^2d_{\alpha\beta}\Big)\varepsilon_\perp^{\beta i}s_{\perp i}
-\lambda f_{L}^\perp\varepsilon_{\perp \alpha i}k_\perp^i+... \ , \label{phi0-lorentz}\\
 \tilde \Phi^{(0)}_{\alpha}=&-\Big(\lambda g_{1L}-\frac{k_\perp\cdot s_\perp}{M}g_{1T}^\perp\Big)p_\alpha
 + g^\perp\varepsilon_{\perp\alpha i}k_\perp^i-g_{T}Ms_{\perp \alpha}\nonumber\\
 &+\frac{g_T^\perp}{M}\Big(k_{\perp\alpha}k_{\perp\beta}-\frac{1}{2}k_\perp^2d_{\alpha\beta}\Big)s_{\perp}^{\beta}-\lambda g_{L}^\perp k_{\perp\alpha}+... \ , \label{phi0t-lorentz}
\end{align}
where $\varepsilon_\perp^{\mu\nu}\equiv\varepsilon^{\rho\sigma\mu\nu}\bar n_\rho n_\sigma$,   $d_{\mu\nu}\equiv g_{\mu\nu}-\bar n_\mu n_\nu-\bar n_\nu n_\mu$, 
and $\varepsilon_\perp^{ks}\equiv({1}/{M})\varepsilon_\perp^{ij}k_{\perp i}s_{\perp j}=({1}/{M})(\vec k_\perp\times\vec s_\perp)\cdot\hat z$. 
\begin{eqnarray}
 &&{\rm Tr}\big[\hat h^{(0)}_{\mu\nu}\slashed p\big]= -4d_{\mu\nu},\\
 &&{\rm Tr}\big[\hat h^{(0)}_{\mu\nu}\gamma_5\slashed p\big]= 4i\varepsilon_{\perp\mu\nu},\\
 &&{\rm Tr}\big[\hat h^{(0)}_{\mu\nu}\gamma_\alpha\big]= \frac{4}{p^+}n_{\{\mu} d_{\nu\}\alpha},\\
 &&{\rm Tr}\big[\hat h^{(0)}_{\mu\nu}\gamma_5\gamma_\alpha\big]= \frac{4i}{p^+}n_{[\mu}\varepsilon_{\perp \nu]\alpha}.
\end{eqnarray}
where
$A_{\{\mu}B_{\nu\}}\equiv A_\mu B_\nu+A_\nu B_\mu$, and $A_{[\mu}B_{\nu]}\equiv A_\mu B_\nu-A_\nu B_\mu$. 
Hence, we obtain, up to twist-3, 
\begin{align}
  \frac{d^2\tilde W^{(0)}_{\mu\nu}}{d^2k_\perp}=&-d_{\mu\nu}\big(f_1-\varepsilon_\perp^{ks}f_{1T}^\perp\big)+\frac{1}{p\cdot q}k_{\perp\{\mu}(q+x_Bp)_{\nu\}}\big(f^\perp-\varepsilon_\perp^{ks}f_{T}^\perp\big)\nonumber\\
  &-\frac{M}{p\cdot q}(q+x_Bp)_{\{\mu}\varepsilon_{\perp \nu\}i}s_\perp^i\hat f_{T}-\frac{\lambda}{p\cdot q}(q+x_Bp)_{\{\mu}\varepsilon_{\perp \nu\}i}k_\perp^if_{L}^\perp\nonumber\\
  &+i\varepsilon_{\perp\mu\nu}\Big(\lambda g_{1L}-\frac{k_\perp\cdot s_\perp}{M}g_{1T}^\perp\Big)-\frac{i}{p\cdot q}k_{\perp[\mu}(q+x_Bp)_{\nu]}\big(g^\perp+\varepsilon_\perp^{ks} g_{T}^\perp\big)\nonumber\\
  &+\frac{iM}{p\cdot q}(q+x_Bp)_{[\mu}\varepsilon_{\perp \nu]i}s_\perp^i \hat g_{T}+\frac{i\lambda}{p\cdot q}(q+x_Bp)_{[\mu}\varepsilon_{\perp\nu]i}k_\perp^ig_{L}^\perp,\label{wmunu-0}
\end{align}
where $\hat f_T=f_T-\frac{k_\perp^2}{2M^2}f_T^\perp$ and $\hat g_T=g_T-\frac{k_\perp^2}{2M^2}g_T^\perp$.

Then, we calculate the contributions from ${d^2\tilde W^{(1)}_{\mu\nu}}/{d^2k_\perp}$. 
Up to twist-3, in the correlation matrix $\varphi^{(1)}_{\rho\alpha}$ and $\tilde \varphi^{(1)}_{\rho\alpha}$,  
we need to consider the $p_\alpha$-terms as given in the following,  
\begin{align}
 \varphi^{(1)}_{\rho\alpha}&= p_\alpha\Big[\varphi^\perp k_{\perp\rho}-\varphi_{T}M\varepsilon_{\perp \rho i}s_\perp^i
 -\frac{\varphi_T^\perp}{M}\Big(k_{\perp\alpha}k_{\perp\beta}-\frac{1}{2}k_\perp^2d_{\alpha\beta}\Big)\varepsilon_\perp^{\beta i}s_{\perp i}
 -\lambda\varphi_{L}^\perp\varepsilon_{\perp \rho i}k_\perp^i\Big]+...\label{phi1-lorentz}\\
 \tilde \varphi^{(1)}_{\rho\alpha}&=ip_\alpha\Big[-\tilde \varphi^{\perp }\varepsilon_{\perp \rho i}k_\perp^i+\tilde \varphi_{T}M s_{\perp\rho}
 -\frac{\tilde\varphi_T^\perp}{M}\Big(k_{\perp\alpha}k_{\perp\beta}-\frac{1}{2}k_\perp^2d_{\alpha\beta}\Big)s_{\perp}^{\beta} 
 +\lambda\tilde \varphi_{L}^{\perp}k_{\perp\rho}\Big]+...
\label{phi1t-lorentz}
\end{align}
The corresponding hard factors are given by,
\begin{eqnarray}
  &&{\rm Tr}\big[\hat h^{(1)\rho}_{\mu\nu}\slashed p\big]=-8 p_\mu{d_{\nu}}^\rho, \\
  &&{\rm Tr}\big[\hat h^{(1)\rho}_{\mu\nu}\gamma_5\slashed p\big]=-8{i}p_\mu{\varepsilon_{\perp\nu}}^\rho.
\end{eqnarray}
We insert them into Eq. \eqref{dwdk-1}, and obtain, 
\begin{align}
  \frac{d^2\tilde W^{(1,L)}_{\mu\nu}}{d^2k_\perp^\prime}=&-\frac{p_\mu}{p\cdot q}\left[\big(\varphi^\perp-\varepsilon_\perp^{ks}\varphi_{T}^\perp\big)k_{\perp\nu}-\hat\varphi_{T}M\varepsilon_{\perp\nu i}s_\perp^i-\lambda\varphi_{L}^\perp\varepsilon_{\perp \nu i}k_\perp^i\right]\nonumber\\
  &-\frac{p_\mu}{p\cdot q}\left[\big(\tilde \varphi^\perp+\varepsilon_\perp^{ks}\tilde\varphi_{T}^\perp\big)k_{\perp\nu}+\hat{\tilde\varphi}_{T}M\varepsilon_{\perp\nu i}s_\perp^i+\lambda\tilde\varphi_{L}^\perp\varepsilon_{\perp \nu i}k_\perp^i\right],
\end{align}
where $\hat \varphi_T=\varphi_T-\frac{k_\perp^2}{2M^2}\varphi_T^\perp$ and $\hat{\tilde\varphi}_T=\tilde\varphi_T-\frac{k_\perp^2}{2M^2}\tilde\varphi_T^\perp$.

\subsection{Simplifying ${d^2W_{\mu\nu}}/{d^2k_\perp}$ with QCD EOM relations}

The quark field operator $\psi (y)$ satisfies the QCD equation of motion (EOM) for massless quark $\gamma\cdot D(y) \psi (y)=0$. 
Hence, the correlation functions defined in 
Eqs. \eqref{phi0-lorentz}, \eqref{phi0t-lorentz}, \eqref{phi1-lorentz}, \eqref{phi1t-lorentz} 
are not independent from each other.
We have in particular, for $\rho=1,2$, 
\begin{align}
x\Phi^{(0)}_{\rho}&=-\frac{n^\alpha}{p^+}\left({\rm Re}\varphi^{(1)}_{\rho\alpha}-{\varepsilon_{\perp\rho}}^\sigma {\rm Im}\tilde\varphi^{(1)}_{\sigma\alpha}\right),\label{eom-1}\\
x\tilde\Phi^{(0)}_{\rho}&=-\frac{n^\alpha}{p^+}\left({\rm Re}\tilde\varphi^{(1)}_{\rho\alpha}+{\varepsilon_{\perp\rho}}^\sigma {\rm Im}\varphi^{(1)}_{\sigma\alpha}\right).\label{eom-2}
\end{align}
We make Lorentz contractions of both sides of Eq. \eqref{eom-1} with $k_\perp^\rho$ and $\varepsilon_\perp^{\rho i}k_{\perp i}$, and obtain, 
\begin{align}
&xf^\perp=-{\rm Re}\big(\varphi^\perp-\tilde \varphi^\perp\big), \label{eom-1a}\\
&xf_{T}=-{\rm Re}\big(\varphi_{T}+\tilde \varphi_{T}\big),\label{eom-1b}\\
&xf_{L}^\perp=-{\rm Re}\big(\varphi_{L}^\perp+\tilde\varphi_{L}^\perp\big),\label{eom-1c}\\
&xf_{T}^\perp=-{\rm Re}\big(\varphi_{T}^\perp+\tilde \varphi_{T}^\perp\big),
\end{align}
Similarly, after Lorentz contractions of both sides of Eq. \eqref{eom-2} with $k_\perp^\rho$ and $\varepsilon_\perp^{\rho i}k_{\perp i}$ , we obtain, 
\begin{align}
&xg^\perp ={\rm Im}\big(\varphi^\perp-\tilde \varphi^\perp\big),\label{eom-2a}\\
&xg_{T}=-{\rm Im}\big(\varphi_{T}+\tilde\varphi_{T} \big),\label{eom-2b}\\
&xg_{L}^\perp=-{\rm Im}\big(\varphi_{L}^\perp+\tilde\varphi_{L}^\perp\big),\label{eom-2c}\\
&xg_{T}^\perp=-{\rm Im}\big(\varphi_{T}^\perp+\tilde\varphi_{T}^\perp\big).\label{eom-2d}
\end{align}
We note that, similar relations have also been obtained earlier in e.g. \cite{Bacchetta:2006tn}.
However, the twist-3 parton correlation functions in the corresponding equations 
in \cite{Bacchetta:2006tn} are defined using the quark-gluon-quark correlater 
where gluon field $A_\rho$ is used instead of $D_\rho$ used here. 
We see clearly the similarities and differences between those relations obtained there 
and those that listed above.
 
Using these relations, we re-write the contributions from $\tilde W^{(1)}_{\mu\nu}$ as,
\begin{eqnarray}
  2{\rm Re}\frac{d^2\tilde W^{(1,L)}_{S,\mu\nu}}{d^2k_\perp} &=& \frac{x_B}{p\cdot q}\Big\{
  p_{\{\mu}k_{\perp \nu\}} \big(f^\perp-\varepsilon_\perp^{ks}f_T^\perp\big) 
  -Mp_{\{\mu}\varepsilon_{\perp\nu\}i}s_\perp^i \hat f_T -
  \lambda p_{\{\mu}\varepsilon_{\perp\nu\}i}k_\perp^i f_L^\perp   \Big\},  \\
  2{\rm Im}\frac{d^2\tilde W^{(1,L)}_{A,\mu\nu}}{d^2k_\perp} &=& \frac{x_B}{p\cdot q}\Big\{ 
    p_{[\mu}k_{\perp \nu]} \big(g^\perp+\varepsilon_\perp^{ks}g_T^\perp\big)
  +Mp_{[\mu}\varepsilon_{\perp\nu]i}s_\perp^i \hat g_T
  +\lambda p_{[\mu}\varepsilon_{\perp\nu]i}k_\perp^ig_L^\perp) \Big\}. 
\end{eqnarray}
It is very interesting to see that, up to twist-3, all the contributions can be expressed 
by the coefficient functions of $\hat\Phi^{(0)}$. 

We add the contributions from $\tilde W_{\mu\nu}^{(1)}$ to those from $\tilde W^{(0)}_{\mu\nu}$, and obtain 
the final result for the hadronic tensor up to twist-3 as,
\begin{align}
  \frac{d^2 W_{\mu\nu}}{d^2k_\perp}&=-d_{\mu\nu}\big(f_1-\varepsilon_\perp^{ks}f_{1T}^\perp\big)+\frac{1}{p\cdot q}k_{\perp\{\mu}(q+2x_Bp)_{\nu\}}\big(f^\perp-\varepsilon_\perp^{ks}f_T^\perp\big)\nonumber\\
  &-\frac{M}{p\cdot q}(q+2x_Bp)_{\{\mu}\varepsilon_{\perp \nu\}i}s_\perp^i\hat f_T-\frac{\lambda}{p\cdot q}(q+2x_Bp)_{\{\mu}\varepsilon_{\perp \nu\}i}k_\perp^if_L^\perp\nonumber\\
  &+i\varepsilon_{\perp\mu\nu}\Big(\lambda g_{1L}-\frac{k_\perp\cdot s_\perp}{M}g_{1T}^\perp\Big)-\frac{i}{p\cdot q}k_{\perp[\mu}(q+2x_Bp)_{\nu]}\big(g^\perp+\varepsilon_\perp^{ks}g_T^\perp\big)\nonumber\\
  &+\frac{iM}{p\cdot q}(q+2x_Bp)_{[\mu}\varepsilon_{\perp \nu]i}s_\perp^i\hat g_T+\frac{i\lambda}{p\cdot q}(q+2x_Bp)_{[\mu}\varepsilon_{\perp\nu]i}k_\perp^ig_L^\perp\label{wmunu-all}.
\end{align}
We see that the result satisfies the electromagnetic gauge invariance 
$q^\mu{d^2 W_{\mu\nu}}/{d^2k_\perp}=q^\nu{d^2 W_{\mu\nu}}/{d^2k_\perp}=0$ explicitly. 
The result is expressed in terms of 12 transverse momentum dependent (TMD) parton distribution and/or correlation functions. 
They contain the information from hadron structure and that from the multiple gluon scattering. We discuss them briefly in the following section. 

\subsection{TMD quark distribution/correlation functions}

As can be seen from Eq. \eqref{wmunu-all}, up to twist-3, 12 TMD parton distribution and/or correlation functions are involved for the semi-inclusive 
DIS scattering process $e^-+N\to e^-+q+X$. 
Six of them are from the expansion of $\Phi_{\alpha}^{(0)}=\rm {Tr}[\gamma_\alpha\hat \Phi_{\alpha}^{(0)}]/2$ and six from 
$\tilde \Phi_{\alpha}^{(0)}=\rm {Tr}[\gamma_5\gamma_\alpha\hat \Phi_{\alpha}^{(0)}]/2$. 
They are defined in Eqs. \eqref{phi0-lorentz} and \eqref{phi0t-lorentz}. 
By reversing these two equations, we can obtain the operator expressions for these quark distribution and correlation functions.
Four of them are leading twist parton distribution functions 
%
are quite familiar with us and can be found 
in different literature, e.g., in~\cite{Boer:1997nt}.  
They all have clear physical interpretations 
and have attracted much attention and  have been with much efforts both theoretically~\cite{Ji:2002aa,Belitsky:2002sm,Ji:2004wu,Ji:2006br,Ji:2006vf,Collins:2011ca,Collins:2011zzd,Aybat:2011ge,Yuan:2003wk,Avakian:2010br,Pasquini:2009bv,Pasquini:2010af,Musch:2009ku,Kotzinian:2006dw,Anselmino:2008sga,Bacchetta:2006tn,Boer:2011fh} 
and experimentally~\cite{Airapetian:2004tw,Airapetian:2009ae,Alexakhin:2005iw,Alekseev:2010rw,Alekseev:2008aa,Alekseev:2010dm,Webb:2005cd,Asaturyan:2011mq,Gao:2010av,Qian:2011py,Huang:2011bc,Pappalardo:2011cu,Avakian:2010ae}.
As can been seen in Sec II.B, in the jet production process $e+N\to e+q+X$ 
where only one hadron state is involved, 
the hard parts contain odd number of $\gamma-$matrices. 
Hence, in the decomposition of correlation matrices, 
chiral-odd distribution and/or correlation functions 
such as transversity and Boer-Mulders functions, involve even number of $\gamma-$matrices 
and will not contribute. 
Such functions can be studied in the hadron production process $e+N\to e+h+x$ or 
the Drell-Yan process $p+p\to l\bar l+X$, where two hadron states are involved. 

The other 8 are twist-3 and have the following operator expressions~\cite{Goeke:2005hb}, 
\begin{align}
  k_\perp^2 f^\perp(x,k_\perp)&=\int \frac{p^+dy^-d^2y_\perp}{2(2\pi)^3}e^{ixp^+y^--i\vec k_\perp\cdot\vec y_\perp}
    \langle p|\bar\psi(0)\slashed k_\perp{\cal L}(0;y)\psi(y)|p\rangle,\label{fperp}\\
  k_\perp^2 g^\perp(x,k_\perp)&=-\int \frac{p^+dy^-d^2y_\perp}{2(2\pi)^3}e^{ixp^+y^--i\vec k_\perp\cdot\vec y_\perp}
    \langle p|\bar\psi(0)\varepsilon_\perp^{ij}k_{\perp j} \gamma_{\perp i}\gamma_5{\cal L}(0;y)\psi(y)|p\rangle,\label{tilde-fperp}\\
  \varepsilon_\perp^{ks} \left(k_\perp\cdot  s_\perp\right)f_T(x,k_\perp)&=-\frac{1}{M^2}\int \frac{p^+dy^-d^2y_\perp}{2(2\pi)^3}e^{ixp^+y^--i\vec k_\perp\cdot\vec y_\perp}\nonumber\\
  &~~\times \langle p,s_\perp^{\uparrow\downarrow}|\bar\psi(0)\big(k_\perp^i k_\perp^j-\frac{1}{2}k_\perp^2 d^{ij}\big)\gamma_{\perp i}s_{\perp j}{\cal L}(0;y)\psi(y)|p,s_\perp^{\uparrow\downarrow}\rangle,\label{f-perp-s1}\\
  \varepsilon_\perp^{ks} \left(k_\perp\cdot  s_\perp\right)g_T(x,k_\perp)&=\frac{1}{M^2}\int \frac{p^+dy^-d^2y_\perp}{2(2\pi)^3}e^{ixp^+y^--i\vec k_\perp\cdot\vec y_\perp}\nonumber\\
  &~~\times \langle p,s_\perp^{\uparrow\downarrow}|\bar\psi(0) \big(k_\perp^i k_\perp^j-\frac{1}{2}k_\perp^2 d^{ij}\big)\gamma_{\perp i}\varepsilon_{\perp jl}s_\perp^l\gamma_5 {\cal L}(0;y)\psi(y)|p,s_\perp^{\uparrow\downarrow}\rangle,\label{tilde-f-perp-s1}\\
  \varepsilon_\perp^{ks} \left(k_\perp\cdot  s_\perp\right)f_T^\perp(x,k_\perp)&=-\int \frac{p^+dy^-d^2y_\perp}{2(2\pi)^3}e^{ixp^+y^--i\vec k_\perp\cdot\vec y_\perp}
    \langle p,s_\perp^{\uparrow\downarrow}|\bar\psi(0)
    \slashed s_\perp
    {\cal L}(0;y)\psi(y)|p,s_\perp^{\uparrow\downarrow}\rangle,\label{f-perp-s3}\\
  \varepsilon_\perp^{ks} \left(k_\perp\cdot  s_\perp\right)g_T^\perp(x,k_\perp)&=-\int \frac{p^+dy^-d^2y_\perp}{2(2\pi)^3}e^{ixp^+y^--i\vec k_\perp\cdot\vec y_\perp}
    \langle p,s_\perp^{\uparrow\downarrow}|\bar\psi(0)
    \varepsilon_\perp^{ij} s_{\perp j} \gamma_{\perp i}\gamma_5
    {\cal L}(0;y)\psi(y)|p,s_\perp^{\uparrow\downarrow}\rangle,\label{tilde-f-perp-s3}\\
  k_\perp^2 f_L^\perp(x,k_\perp)&=-\int \frac{p^+dy^-d^2y_\perp}{2(2\pi)^3}e^{ixp^+y^--i\vec k_\perp\cdot\vec y_\perp}
    \langle p,+|\bar\psi(0)\varepsilon_\perp^{ij} k_{\perp j} \gamma_{\perp i}{\cal L}(0;y)\psi(y)|p,+\rangle,\label{f-perp-s2}\\
  k_\perp^2 g_L^\perp(x,k_\perp)&=\int \frac{p^+dy^-d^2y_\perp}{2(2\pi)^3}e^{ixp^+y^--i\vec k_\perp\cdot\vec y_\perp}
    \langle p,+|\bar\psi(0)\slashed k_\perp\gamma_5{\cal L}(0;y)\psi(y)|p,+\rangle. \label{tilde-f-perp-s2}
\end{align}
Among them, $f^\perp$ and $g^\perp$ are related to the unpolarized case; 
$f_T$, $g_T$, $f_T^\perp$, and $g_T^\perp$ are related to the transverse polarization 
and $f_L^\perp$ and $g_L^\perp$  are related to the longitudinal polarization. 

These twist-3 quark correlation functions have no simple probabilistic interpretation. 
In fact, as we can see from the derivations that lead to these results, these twist-3 correlation functions   
come from the interference terms between amplitudes for scattering without multiple gluon scattering and that with one gluon scattering.  

If we integrate over  $\int d^2k_\perp$, we obtain the hadronic tensor $W_{\mu\nu}$ from $d^2W_{\mu\nu}/d^2k_\perp$. 
Since all parton distribution and/or correlation functions $f$'s and $g$'s are scalar functions of $k_\perp$, 
all the terms that are linearly dependent on $k_\perp$ vanish after the integration and we obtain from Eq.~\eqref{wmunu-all} that,  
\begin{align}
  W_{\mu\nu}=&-d_{\mu\nu}f_1(x)-\frac{M}{p\cdot q}(q+2x_Bp)_{\{\mu}\varepsilon_{\perp \nu\}i}s_\perp^if_T(x)\nonumber\\
  &+i\varepsilon_{\perp\mu\nu}\lambda g_{1L}(x)+\frac{iM}{2p\cdot q}(q+2x_Bp)_{[\mu}\varepsilon_{\perp \nu]i}s_\perp^ig_T(x) ,\label{w_integrate}
\end{align}
where $f_1(x)\equiv \int d^2k_\perp f_1(x,k_\perp)$, $g_{1L}(x)\equiv \int d^2k_\perp g_{1L}(x,k_\perp)$, and,
\begin{align}
f_T(x)&\equiv\int d^2k_\perp f_T(x,k_\perp),\label{f-T-prime}\\
g_T(x)&\equiv\int d^2k_\perp g_T(x,k_\perp).\label{g-T-prime}
\end{align}
The $g_T(x)$ term is the only twist-3 contribution to the structure function in inclusive DIS 
with longitudinally polarized lepton beam and transversely polarized nucleon target, as 
discussed in~\cite{Ji:1993ey}. 
The $f_T(x)$ term is a time-reversal-odd term corresponding to the T-odd term 
$p_{\{\mu}\varepsilon_{\nu\}\rho\sigma\tau}p^\rho q^\sigma s^\tau$ in $W_{\mu\nu}$. 
It can be shown that,  under time reversal invariance, $f_T(x)=0$. 

The situation as considered by in~\cite{Cahn:1978se} and~\cite{Liang:1993re} can be recovered by putting $g=0$. 
In this case, there is no multiple gluon scattering and ${\cal L}=1$. 
Consequently the T-odd TMD distribution and/or correlation functions must be zero. 
The twist-3 quark correlation functions reduces to,
\begin{align}
  &xf^\perp(x,k_\perp)\big|_{g=0}=f_1(x,k_\perp)\big|_{g=0},                                                                   \label{gto0-a}\\
  &f_T(x,k_\perp)\big|_{g=0}=f_L^\perp(x,k_\perp)\big|_{g=0}=f_T^\perp(x,k_\perp)\big|_{g=0}=0,                \label{gto0-b}\\
  &xg_L^\perp(x,k_\perp)\big|_{g=0}=g_{1L}(x,k_\perp)\big|_{g=0},                                                         \label{gto0-c}\\
  &xg_T(x,k_\perp)\big|_{g=0}=-\frac{k_\perp^2}{2M^2}xg_T^\perp(x,k_\perp)\big|_{g=0}=-\frac{k_\perp^2}{2M^2}g_{1T}^\perp(x,k_\perp)\big|_{g=0}                                    \label{gto0-d}\\
  &g^\perp(x,k_\perp)\big|_{g=0}=0.       \label{gto0-e}
\end{align}
The hadronic tensor reduces to,
\begin{align}
 \frac{d^2\tilde W_{\mu\nu}}{d^2k_\perp}\big|_{g=0}=&
  -\Big[d_{\mu\nu}-\frac{1}{x_Bp\cdot q}k_{\perp\{\mu}(q+2x_Bp)_{\nu\}}\Big]f_1(x,k_\perp)\big|_{g=0} \nonumber\\
  &+i\lambda\Big[\varepsilon_{\perp\mu\nu}+\frac{1}{x_Bp\cdot q}(q+2x_Bp)_{[\mu}\varepsilon_{\perp \nu]i}k_\perp^i\Big]g_{1L}(x,k_\perp)\big|_{g=0} \nonumber\\
  &-i\frac{k_\perp\cdot s_\perp}{M}\Big[\varepsilon_{\perp\mu\nu}+\frac{1}{x_Bp\cdot q}(q+2x_Bp)_{[\mu}\varepsilon_{\perp\nu]i}k_\perp^i\Big]g_{1T}^\perp(x,k_\perp)\big|_{g=0}.
\end{align}
This just corresponds to the results obtained using the simple parton model with intrinsic transverse momentum as discussed 
in~\cite{Cahn:1978se} and~\cite{Liang:1993re} for the unpolarized and the longitudinally polarized case respectively. 
The deviations from this result come from the multiple gluon scattering.

\section{Cross sections and azimuthal asymmetries}

Making the Lorentz contraction of the hadronic tensor $d^2W_{\mu\nu}/d^2k_\perp$ as given by Eq. \eqref{wmunu-all} with the leptonic tensor $L^{\mu\nu}(l,l')$,
we obtain the differential cross section of the process $e^-(l,s_l)+N(p,s)\to e^-(l^\prime)+q(k')+X$ as, 
\begin{equation}
  \frac{d\sigma}{dx_Bdyd^2k_\perp}=\frac{2\pi\alpha_{\rm em}^2e_q^2}{Q^2y}\left(W_{UU}+\lambda_lW_{LU}+s_\perp W_{UT}+\lambda W_{UL}+\lambda_l\lambda W_{LL}+\lambda_l s_\perp W_{LT}\right),
 \end{equation}
where $W_{s_ls}$ represents the contribution in the different polarization case,  and we use 
the superscript $s_l=U$ or $L$ to denote whether the lepton is unpolarized or longitudinally polarized, 
while $s=U, L$ or $T$ denotes whether the nucleon is unpolarized, longitudinally 
or transversely polarized~\cite{ff1}. 
These different contributions are given by,
\begin{align}
  W_{UU}=&A(y)f_1-\frac{2x_B  |\vec  k_\perp|}{Q}B(y)f^\perp\cos\phi,\label{F-UU}\\
  W_{UT}=&\frac{ |\vec  k_\perp|}{M}A(y) f_{1T}^{\perp}\sin\left(\phi-\phi_s\right)\nonumber\\
  &- \frac{2x_BM}{Q} B(y) \Big[ f_T\sin\phi_s-\frac{k_\perp^2}{2M^2}f_T^\perp\sin(2\phi-\phi_s)\Big],\label{F-UT}\\
  W_{UL}=&-\frac{ 2x_B|\vec k_\perp|}{Q}B(y)f_L^\perp\sin\phi,\label{F-UL}\\
  W_{LU}=&-\frac{ 2x_B|\vec k_\perp|}{Q}D(y)g^\perp\sin\phi,\label{F-LU}\\
  W_{LL}=& C(y)g_{1L} - \frac{2x_B |\vec k_\perp|}{Q}D(y)g_L^\perp\cos\phi,\label{F-LL}\\
  W_{LT}=&\frac{|\vec k_\perp|}{M}C(y)g_{1T}^\perp \cos\left(\phi-\phi_s\right) \nonumber\\
   & - \frac{2x_BM}{Q} D(y) \Big[ g_T\cos\phi_s-\frac{k_\perp^2}{2M^2}g_T^\perp\cos\left(2\phi-\phi_s\right) \Big] .\label{F-LT}
\end{align}
where $A(y)=1+(1-y)^2$, $B(y)=2(2-y)\sqrt{1-y}$, $C(y)=y(2-y)$, $D(y)=2y\sqrt{1-y}$, 
$\cos\phi={\vec l_\perp\cdot \vec k_\perp}/{|\vec l_\perp||\vec k_\perp|}$, 
$\sin\phi={(\vec l_\perp\times \vec k_\perp)\cdot \vec e_z}/{|\vec l_\perp||\vec k_\perp|}$, 
$\cos\phi_s={\vec l_\perp\cdot \vec s_\perp}/{|\vec l_\perp||\vec s_\perp|}$,
and $\sin\phi_s={(\vec l_\perp\times \vec s_\perp)\cdot\vec e_z}/{|\vec l_\perp||\vec s_\perp|}$. 

We note that, except the slightly different notations~\cite{ff1}, these results are almost
the same as those obtained in \cite{Bacchetta:2006tn} for jet production. 
They have the same structures and the forms of the coefficients in each term are the same~\cite{ff2}.   
This is expected since the kinematics is the same and the approximations made 
in the hard parts in \cite{Bacchetta:2006tn} should be equivalent to keep the leading 
and sub-leading terms in the collinear expansion. 
Also, due to the relationship given by Eqs.~(\ref{eom-2a}-\ref{eom-2d}) obtained from equation of motion, 
all the results are expressed by the different components of $\hat \Phi^{(0)}$ defined by 
Eq. (\ref{eq:Phi0def}) which is identical to the original $\hat \phi^{(0)}$ given 
by Eq. (\ref{eq:phi0}) except for the gauge link.
Hence, the difference in defining higher twist correlators such as that between 
 $\hat \varphi^{(1)}$ given by Eq. (\ref{eq:varphi1}) and $\hat \phi^{(1)}$ given 
by Eq. (\ref{eq:phi1}) does not show up in the final results. 
However, it seems not the case for even higher twist~\cite{Song:2010pf}.  
Other features of the results are summarized in the following.

We see that there are leading twist contributions in the $UU$, $UT$, $LL$ and $LT$ cases, 
while there are twist-3 contributions in all cases.
The different azimuthal asymmetries are defined by the average values of the corresponding 
sine or cosine of the angles. 
There are two leading twist azimuthal asymmetries as given by,
\begin{align}
\langle\sin(\phi-\phi_s)\rangle_{UT}&=s_\perp\frac{ |\vec  k_\perp|}{2M}\frac{f_{1T}^{\perp}(x,k_\perp)}{f_1(x,k_\perp)}, \label{aa:sinphi-phis} \\
\langle\cos(\phi-\phi_s)\rangle_{LT}&=\lambda_ls_\perp\frac{ |\vec  k_\perp|}{2M}\frac{C(y)}{A(y)}\frac{g^\perp_{1T}(x,k_\perp)}{f_1(x,k_\perp)}. \label{aa:cosphi-phis} 
\end{align}
The azimuthal asymmetry $\langle \cos\phi\rangle$ exists at twist-3 for the unpolarized case. 
It receives also a twist-3 contribution in the $LL$ case but also a leading twist contribution in the $LT$ case, i.e., 
\begin{align}
 \langle\cos\phi\rangle_{UU}&=-\frac{|\vec  k_\perp|}{Q}\frac{B(y)}{A(y)}\frac{x_Bf^\perp(x_B,k_\perp)}{f_1(x_B,k_\perp)}, \label{aa:cosphiUU} \\
 \langle\cos\phi\rangle_{LL}&=-\frac{ |\vec  k_\perp|}{Q}\frac{B(y)x_Bf^\perp(x_B,k_\perp)+\lambda_l\lambda D(y)x_Bg^\perp_L(x_B,k_\perp)}
                                                                                              {A(y)f_1(x_B,k_\perp)+\lambda_l\lambda C(y)g_{1L}(x_B,k_\perp)}, \label{aa:cosphiLL} \\
 \langle\cos\phi\rangle_{LT}&=\frac{ |\vec  k_\perp|}{2M} \frac{\lambda_l s_\perp C(y)g_{1T}^\perp(x_B,k_\perp)\cos\phi_s -\frac{2M}{Q}B(y)x_Bf^\perp(x_B,k_\perp)}
 {A(y)f_1(x_B,k_\perp)-\lambda_l s_\perp\frac{2M}{Q}x_Bg_T(x_B,k_\perp)\cos\phi_s}.  \label{aa:cosphiLT}
\end{align}
We note in particular that there exist a twist-3 asymmetry $\langle \sin\phi\rangle$ for the $LU$ or $UL$ case, i.e. 
when lepton or nucleon is longitudinally polarized while the other is unpolarized. 
It is given by,
 \begin{align}
\langle\sin\phi\rangle_{LU} &=
-\lambda_l\frac{|\vec k_\perp|}{Q}\frac{D(y)}{A(y)} \frac{x_Bg^\perp(x_B,k_\perp)}{f_1(x_B,k_\perp)},  \label{aa:sinphiLU} \\
\langle\sin\phi\rangle_{UL} &=
-\lambda \frac{|\vec k_\perp|}{Q}\frac{B(y)}{A(y)}\frac{x_Bf^\perp_L(x_B,k_\perp)}{f_1(x_B,k_\perp)}.  \label{aa:sinphiUL}
\end{align}
They are determined by the TMD parton correlation $g^\perp$ and $f^\perp_T$ respectively. 

It is also interesting to see that, if we integrate over $\phi$, we obtain, 
\begin{align}
 \frac{d\sigma}{|\vec k_\perp| dx_Bdyd|\vec k_\perp|}=&\frac{4\pi^2\alpha_{\rm em}^2e_q^2}{Q^2y} 
         \Big\{  A(y)f_1 -s_\perp \frac{2x_BM}{Q} B(y)  f_T \sin\phi_s \\ 
             &  +\lambda_l\lambda C(y)g_{1L} 
                  -\lambda_ls_\perp \frac{2x_BM}{Q} D(y) g_T\cos\phi_s   \Big\} .
\end{align}
We see that the transverse spin asymmetry exists for the semi-inclusive process $e^-+N\to e^-+q+X$ at the twist-3 level 
both in the target singly polarized case $UT$ and in the case $LT$ where the lepton is also longitudinally polarized. 
But the asymmetries in these two cases are different and are given by, 
\begin{align}
\langle\sin\phi_s\rangle_{UT} &=
-s_\perp\frac{M}{Q}\frac{B(y)}{A(y)}\frac{ x_Bf_T(x_B,k_\perp)}{f_1(x_B,k_\perp)},  \label{aa:sinphisUT} \\
\langle\cos\phi_s\rangle_{LT} &=
-\lambda_l s_\perp\frac{M}{Q}\frac{D(y)}{A(y)}\frac{x_Bg_T(x_B,k_\perp)}{f_1(x_B,k_\perp)}.  \label{aa:sinphisLT}
\end{align}

We also note that, in experiments, we usually measure for a given $|\vec k_\perp|$ interval. In this case, we need to carry out the integration over $|\vec k_\perp|$. 
For example, if we integrate over the whole $|\vec k_\perp|$ region, we obtain, 
\begin{align}
  \langle\langle\sin\phi\rangle\rangle_{LU}=-\lambda_l\frac{B(y)}{A(y)}\frac{2\pi}{Q}\frac{\int \vec k_\perp^2 d|\vec k_\perp|  x_Bg^\perp(x_B,k_\perp)}{f_1(x_B)} .
\end{align}

By carrying out the integration over $d^2k_\perp$, we obtain the differential cross section $d\sigma/dx_Bdy$ for the inclusive DIS process $e^-+N\to e^-+X$ as, 
\begin{align}
  \frac{d\sigma}{dx_Bdy}=&\frac{2\pi\alpha_{\rm em}^2e_q^2}{Q^2y}\Big\{
  A(y)f_1(x_B)+\lambda\lambda_l C(y)g_{1L}(x_B)
  -\lambda_ls_\perp \frac{2x_BM}{Q} D(y)g_T^\prime(x_B)\cos\phi_s \Big\},
\end{align}
where we see clearly that the only twist-3 contribution exists for the case that the lepton is longitudinally polarized and the nucleon is transversely polarized.

At $g=0$, the cross section reduces to the result obtained in the simple parton model with intrinsic transverse momentum~\cite{Cahn:1978se,Liang:1993re}. 
By inserting the results given by Eqs. \eqref{gto0-a} - \eqref{gto0-e} into Eqs. \eqref{F-UU} - \eqref{F-LT}, we obtain, 
\begin{align}
  W_{UU}\big|_{g=0}=&\Big[A(y)-\frac{2|\vec  k_\perp|}{Q}B(y)\cos\phi\Big]f_1(x,k_\perp)\big|_{g=0},\\
  W_{UT}\big|_{g=0}=&F_{UL}\big|_{g=0}=F_{LU}\big|_{g=0}=0, \\
  W_{LL}\big|_{g=0}=&\Big[C(y)-\frac{2 |\vec k_\perp|}{Q}D(y)\cos\phi\Big]g_{1L}(x,k_\perp)\big|_{g=0},\\
  W_{LT}\big|_{g=0}=&  \frac{|\vec k_\perp|}{M}\Big[C(y)-
  \frac{2|\vec k_\perp|}{Q}D(y)\cos\phi\Big]g_{1T}^\perp(x,k_\perp)\big|_{g=0}\cos(\phi-\phi_s) . 
\end{align}
Correspondingly, for the azimuthal asymmetries discussed above, we obtain,
\begin{align}
&\langle\sin(\phi-\phi_s)\rangle_{UT}\big|_{g=0}=0, \label{aa:sinphiphisg=0}\\
&\langle\cos(\phi-\phi_s)\rangle_{LT}\big|_{g=0}=\lambda_ls_\perp\frac{ |\vec  k_\perp|}{2M}\frac{C(y)}{A(y)}\frac{g^\perp_{1T}(x,k_\perp)}{f_1(x,k_\perp)}, \\
&\langle\cos\phi\rangle_{UU} \big|_{g=0}=-\frac{B(y)}{A(y)}\frac{|\vec  k_\perp|}{Q},  \\
&\langle\cos\phi\rangle_{LL} \big|_{g=0}=-\frac{ |\vec  k_\perp|}{Q}\frac{B(y)f_1(x,k_\perp)+\lambda_l\lambda D(y)g_{1L}(x,k_\perp)}
                                                                                                       {A(y)f_1(x,k_\perp)+\lambda_l\lambda C(y)g_{1L}(x,k_\perp)},  \\
&\langle\cos\phi\rangle_{LT}\big|_{g=0}=\frac{ |\vec  k_\perp|}{2M} \frac{\lambda_l s_\perp C(y)g_{1T}^\perp(x_B,k_\perp)\cos\phi_s -\frac{2M}{Q}B(y)x_Bf_1(x_B,k_\perp)}
 {A(y)f_1(x_B,k_\perp)+\lambda_l s_\perp\frac{k_\perp^2}{MQ}D(y)g_{1T}^\perp(x,k_\perp)\cos\phi_s},\\
&\langle\sin\phi\rangle_{LU}\big|_{g=0}=\langle\sin\phi\rangle_{UL}\big|_{g=0} =\langle\sin\phi_s\rangle_{UT}\big|_{g=0} =0,\\
&\langle\cos\phi_s\rangle_{LT}\big|_{g=0}=\lambda_l s_\perp\frac{D(y)}{A(y)}\frac{k_\perp^2}{MQ}\frac{g_{1T}^\perp(x_B,k_\perp)}{f_1(x_B,k_\perp)}. \label{aa:cosphisg=0}
\end{align}
Clearly, a systematic study of these asymmetries should provide very important information on the structure of the nucleon and the properties of strong interaction. 
In particular, the deviations from the results given by Eqs. \eqref{aa:sinphiphisg=0} - \eqref{aa:cosphisg=0} tell us the influences from the multiple gluon scattering. 

\section{Nuclear dependence}

The above mentioned calculations apply to $e^-+N\to e^-+q+X$ as well as
$e^-+A\to e^-+q+X$, i.e. for reactions using a nucleus target.
Similar results are obtained with only a replacement of  the state $|N\rangle$ by $|A\rangle$ in the definitions of
the parton distribution and/or correlation functions.
It has also been shown~\cite{Liang:2008vz} that the multiple gluon scattering contained in the gauge link
leads to a strong nuclear dependence for these TMD parton distribution and/or correlation functions.
Such nuclear dependences can manifest themselves in
the azimuthal asymmetries in SIDIS~\cite{Gao:2010mj,Song:2010pf}.
In this section, we present the results for the parton distributions and azimuthal asymmetries given in last section.

\subsection{$A$-dependence of the parton correlation functions}

If we replace the state $|N\rangle$ by $|A\rangle$,  the multiple gluon scattering
in the gauge link can be connected to different nucleons in the nucleus $A$ thus gives rise to nuclear dependence.
It has been shown that,  under the ``maximal two gluon approximation"~\cite{Liang:2008vz},
a TMD quark distribution $\Phi^A_\alpha(x,k_\perp)$ in nucleus defined in the form,
\begin{equation}
\Phi^A_\alpha(x,k_\perp)\equiv \int \frac{p^+dy^-d^2y_\perp}{(2\pi)^3}
e^{ixp^+y^- -i\vec  k_\perp\cdot \vec y_\perp}
\langle A \mid \bar\psi(0)\Gamma_\alpha{\cal L}(0;y)\psi(y)\mid A \rangle,
\label{form}
\end{equation}
is given by a convolution of the corresponding distribution $\Phi^N_\alpha(x,k_\perp)$ in nucleon and a Gaussian broadening~\cite{Liang:2008vz}, i.e., 
\begin{equation}
\Phi^A_\alpha(x,k_\perp)\approx\frac{A}{\pi \Delta_{2F}}
\int d^2\ell_\perp e^{-(\vec k_\perp -\vec\ell_\perp)^2/\Delta_{2F}}\Phi^N_\alpha(x,\ell_\perp),
\label{tmdgeneral}
\end{equation}
where $\Gamma_\alpha$ is any gamma matrix, 
$\Delta_{2F}$ is the broadening width,
$\Delta_{2F}=\int d\xi^-_N \hat q_F(\xi_N)$, 
and $\hat q_F(\xi_N)=(2\pi^2\alpha_s/N_c)\rho_N^A(\xi_N)[xf^N_g(x)]_{x=0}$ is the quark transport parameter,
where $\rho_N^A(\xi_N)$ is the spatial nucleon number density inside the nucleus
and $f^N_g(x)$ is the gluon distribution function in nucleon,
the superscript $A$ or $N$ denotes that it is for the nucleus or the nucleon.

The derivations in~\cite{Liang:2008vz} apply to any nucleon and nucleus in the unpolarized case.
Since both $\Phi_\alpha^{(0)}$ and $\tilde\Phi_\alpha^{(0)}$ defined in Eqs. \eqref{phi0-lorentz} and \eqref{phi0t-lorentz} are 
of the form given by Eq. (\ref{tmdgeneral}), Eq. \ref{tmdgeneral} applies and 
derive the $A$-dependences of different parton distribution and/or correlation functions in the unpolarized case. 
For those involved in the differential cross section up to twist-3, we obtain, 
\begin{align}
  f_1^A(x,k_\perp)&\approx\frac{A}{\pi\Delta_{2F}}\int d^2\ell_\perp e^{-(\vec k_\perp -\vec\ell_\perp)^2/\Delta_{2F}}f_1^N(x,\ell_\perp),\\
  |\vec k_\perp|^2f^{\perp A}(x,k_\perp)&\approx\frac{A}{\pi\Delta_{2F}}\int d^2\ell_\perp e^{-(\vec k_\perp -\vec\ell_\perp)^2/\Delta_{2F}}(\vec k_\perp\cdot \vec \ell_\perp)f^{\perp N}(x,\ell_\perp),\\
  |\vec k_\perp|^2g^{\perp A}(x,k_\perp)&\approx\frac{A}{\pi\Delta_{2F}}\int d^2\ell_\perp e^{-(\vec k_\perp -\vec\ell_\perp)^2/\Delta_{2F}}(\vec k_\perp\cdot\vec \ell_\perp)g^{\perp N}(x,\ell_\perp).
\end{align}

To illustrate the dependence more clearly, we take the Gaussian ansatz for the transverse momentum dependence, i.e., 
\begin{align}
  &f_1^N(x,\ell_\perp)=\frac{1}{\pi\alpha}f_1^N(x)e^{-\vec \ell_\perp^2/\alpha},&\\
  &f^{\perp N}(x,\ell_\perp)=\frac{1}{\pi\beta}f^{\perp N}(x)e^{-\vec \ell_\perp^2/\beta},&\\
  &g^{\perp N}(x,\ell_\perp)=\frac{1}{\pi\gamma}g^{\perp N}(x)e^{-\vec \ell_\perp^2/\gamma},
\end{align}
and obtain immediately, 
\begin{align}
  &f_1^A(x,k_\perp)\approx\frac{A}{\pi\alpha_A}f_1^N(x)e^{-\vec k_\perp^2/\alpha_A},&\\
  &f^{\perp A}(x,k_\perp)\approx\frac{A}{\pi\beta_A}\frac{\beta}{\beta_A}f^{\perp N}(x)e^{-\vec k_\perp^2/\beta_A},&\\
  &g^{\perp A}(x,k_\perp)\approx\frac{A}{\pi\gamma_A}\frac{\gamma}{\gamma_A}g^{\perp N}(x)e^{-\vec k_\perp^2/\gamma_A},\label{nuclearPDF:unpolarized}
\end{align}
where $\alpha_A=\alpha+\Delta_{2F}$, $\beta_A=\beta+\Delta_{2F}$ and $\gamma_A=\gamma+\Delta_{2F}$. 
We see that all the TMD distribution/correlation functions have $p_T$-broadening with the magnitude $\Delta_{2F}$, 
but the twist-3 parton correlation function $f^\perp(x,k_\perp)$ or $g^\perp(x,k_\perp)$ has an extra suppression factor 
$\beta/\beta_A$ or $\gamma/\gamma_A$.

\subsection{$A$-dependence of the azimuthal asymmetry}

Having the nuclear dependences of the TMD parton distribution and correlation functions and the 
expressions for the azimuthal asymmetries presented in the previous sections, 
we can calculate the nuclear dependence of the azimuthal asymmetries in a straight forward manner with the Gaussian ansatz for
the TMD distributions and/or correlations. 

For reactions with unpolarized target, the results are just the same as those for the unpolarized reaction as discussed in~\cite{Gao:2010mj}. 
This applies to the asymmetry $\langle\sin\phi\rangle_{LU}$ given by Eq. \eqref{aa:sinphiLU}, for which we obtain, 
\begin{align}
  \frac{\langle\sin\phi\rangle_{LU}^{eA}}{\langle\sin\phi\rangle_{LU}^{eN}}\approx&\frac{\alpha_A}{\alpha}\Big(\frac{\gamma}{\gamma_A}\Big)^2
  \exp\Big[\Big(\frac{1}{\alpha_A}-\frac{1}{\alpha}-\frac{1}{\gamma_A}+\frac{1}{\gamma}\Big)\vec k_\perp^2\Big].
\end{align}
For $\alpha=\gamma$, it simply reduces to
\begin{align}
  \frac{\langle\sin\phi\rangle_{LU}^{eA}}{\langle\sin\phi\rangle_{LU}^{eN}}\approx\frac{\alpha}{\alpha+\Delta_{2F}}.
\end{align}
Integrated over $|\vec k_\perp|$, we have, 
\begin{align}
  \frac{\langle\langle\sin\phi\rangle\rangle_{LU}^{eA}}{\langle\langle\sin\phi\rangle\rangle_{LU}^{eN}}\approx&\sqrt{\frac{\gamma}{\gamma+\Delta_{2F}}},
\end{align}
We see that, also in this case, the asymmetry is suppressed in reactions using the nucleus target in similar manner as 
in the unpolarized case discussed in~\cite{Gao:2010mj,Song:2010pf}.

The width $\gamma$ can in general be different from $\alpha$. 
Hence, we present as an example in Figs. (1a) and (1b) the ratio as a function of $k_T$-broadening parameter $\Delta_{2F}$. 
\begin{figure}[h!]
 \centering
  \epsfig{file=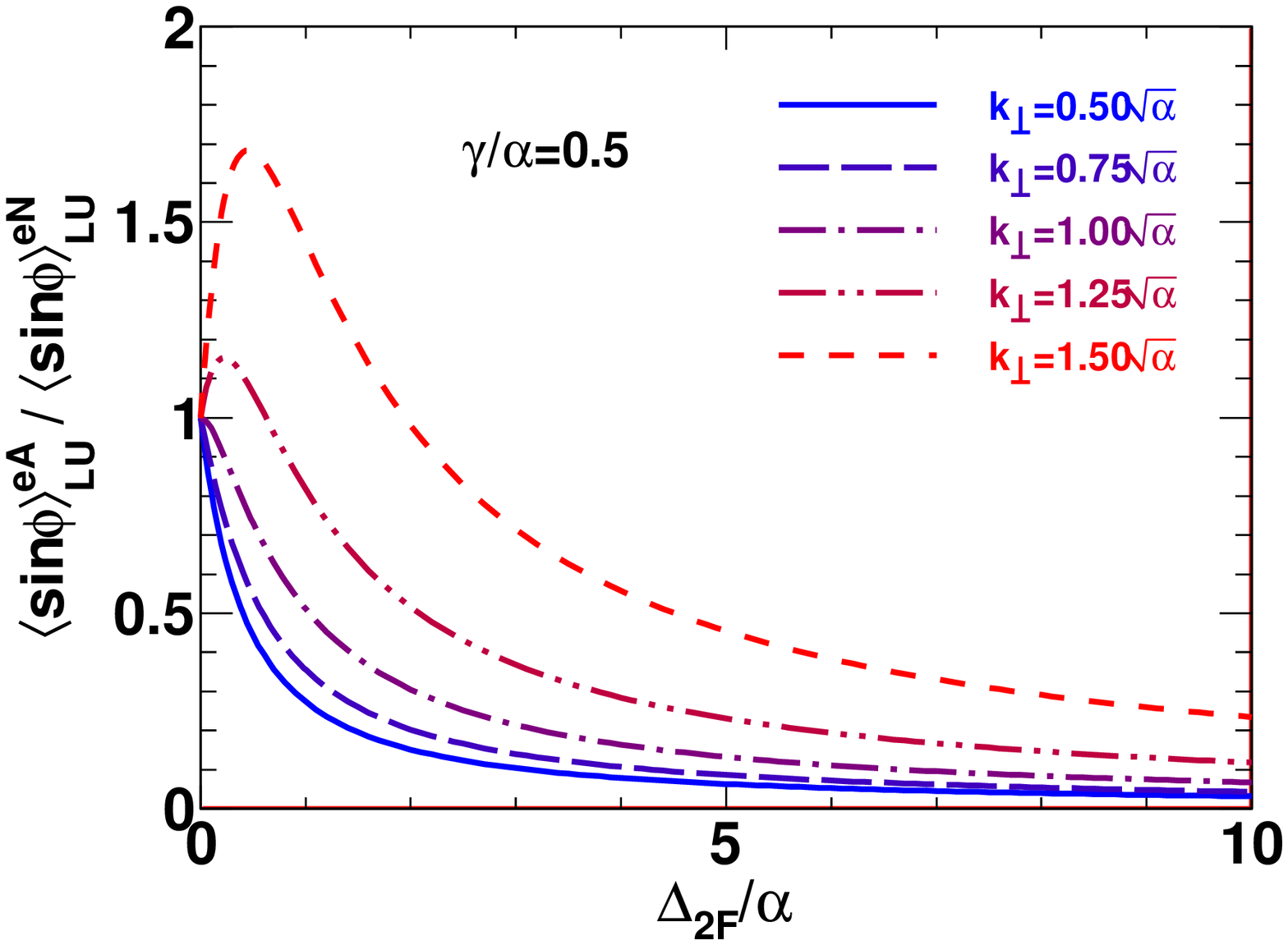,width=0.55\textwidth,clip=}
  \epsfig{file=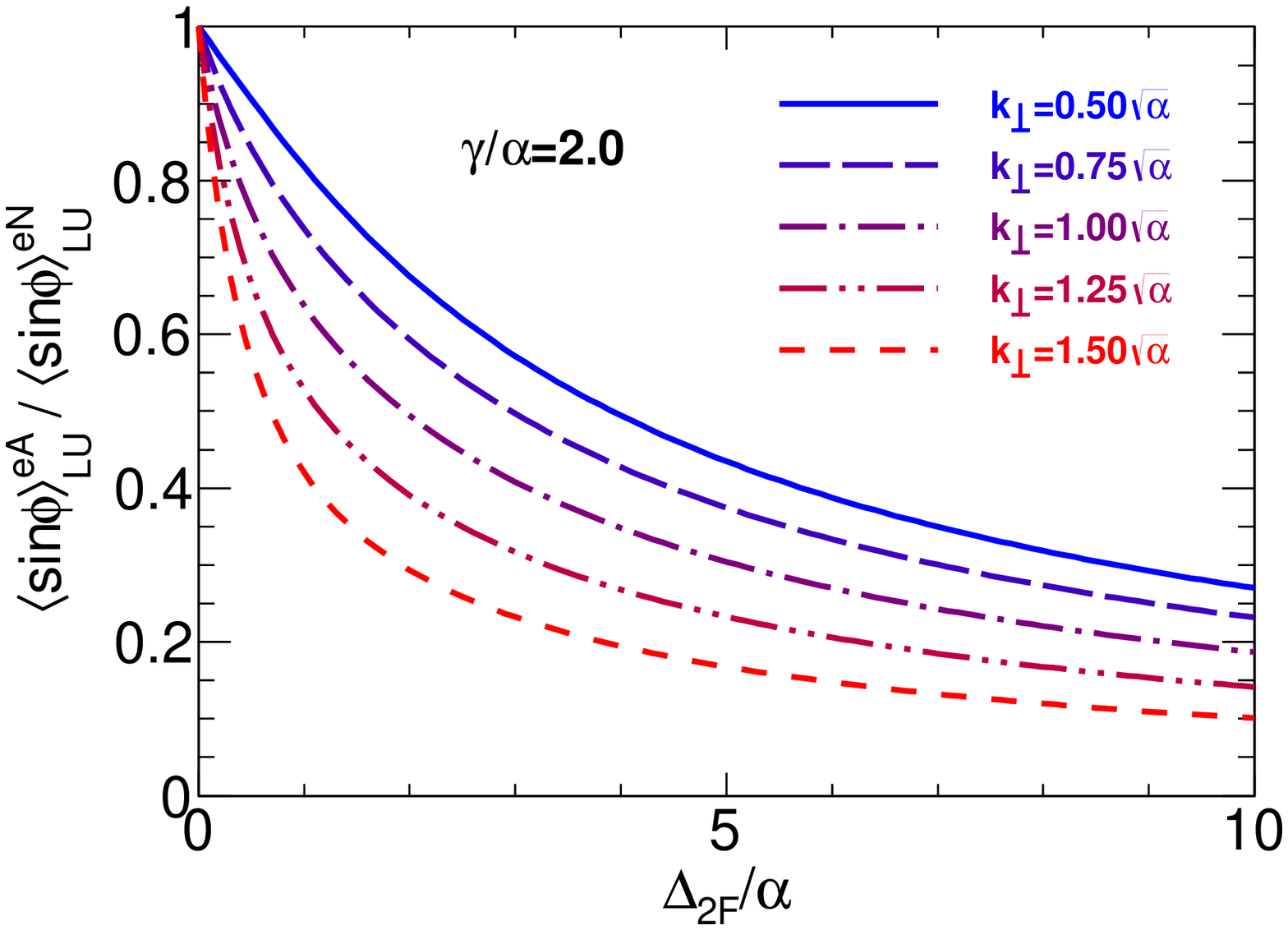,width=0.55\textwidth,clip=}
 \caption{Ratio of $\langle\sin\phi\rangle_{LU}^{eA}/\langle\sin\phi\rangle_{LU}^{eN}$ as a function of $\Delta_{2F}$ for different $k_\perp$ and $\gamma$.}
\end{figure}
We see that it is very similar to $\langle \cos\phi\rangle_{UU}$ discussed in~\cite{Gao:2010mj}. 
We also plot the $k_T$-dependence of the ratio in Figs. (2a) and (2b). 
It is easy to see that for $\gamma/\alpha<1$, the ratio of $\langle\sin\phi\rangle_{LU}$ is quite sensitive to the value of $\gamma/\alpha$ in the large $k_\perp$ region. 
\begin{figure}[h!]
 \centering
  \epsfig{file=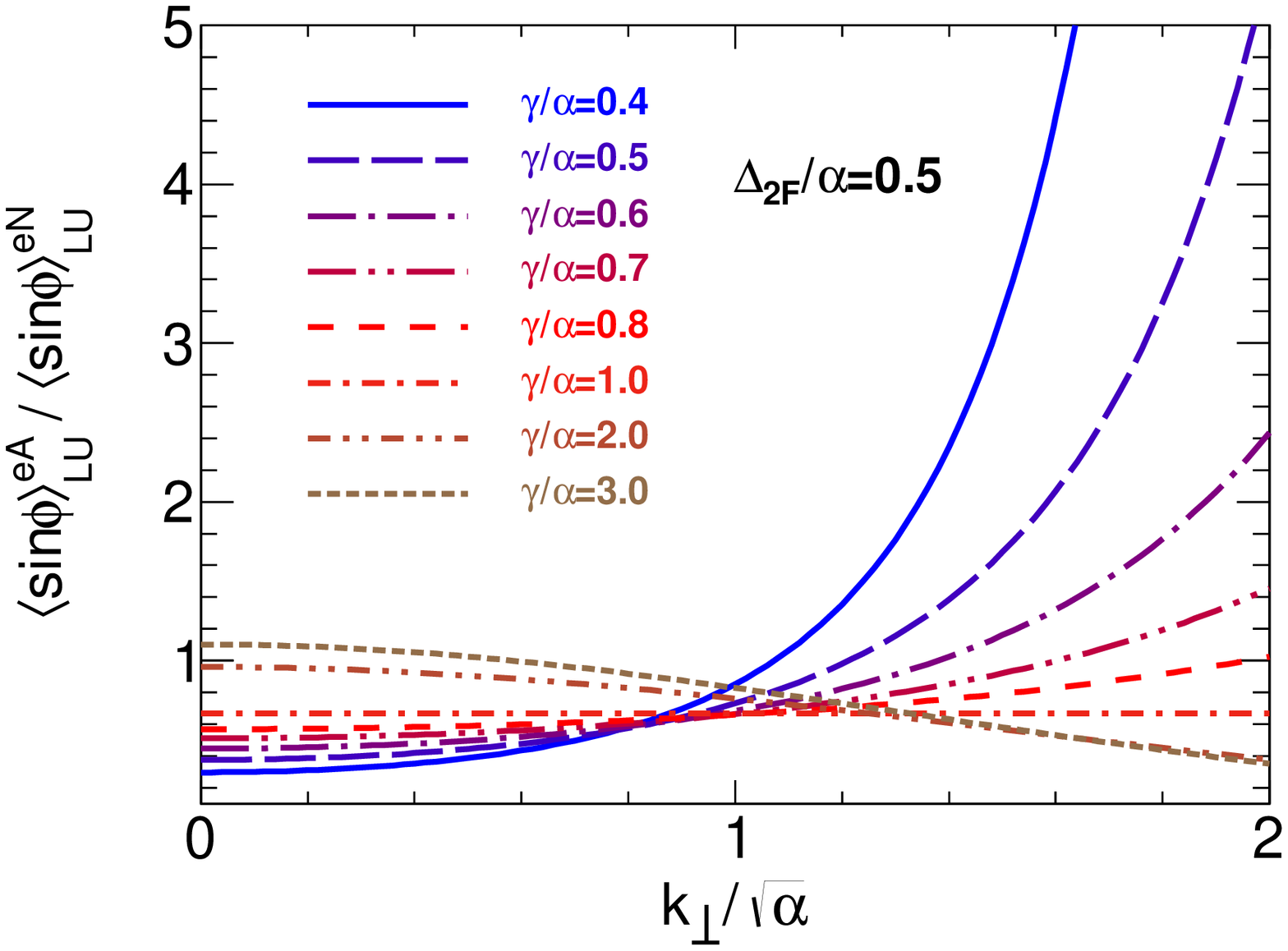,width=0.55\textwidth,clip=}
  \epsfig{file=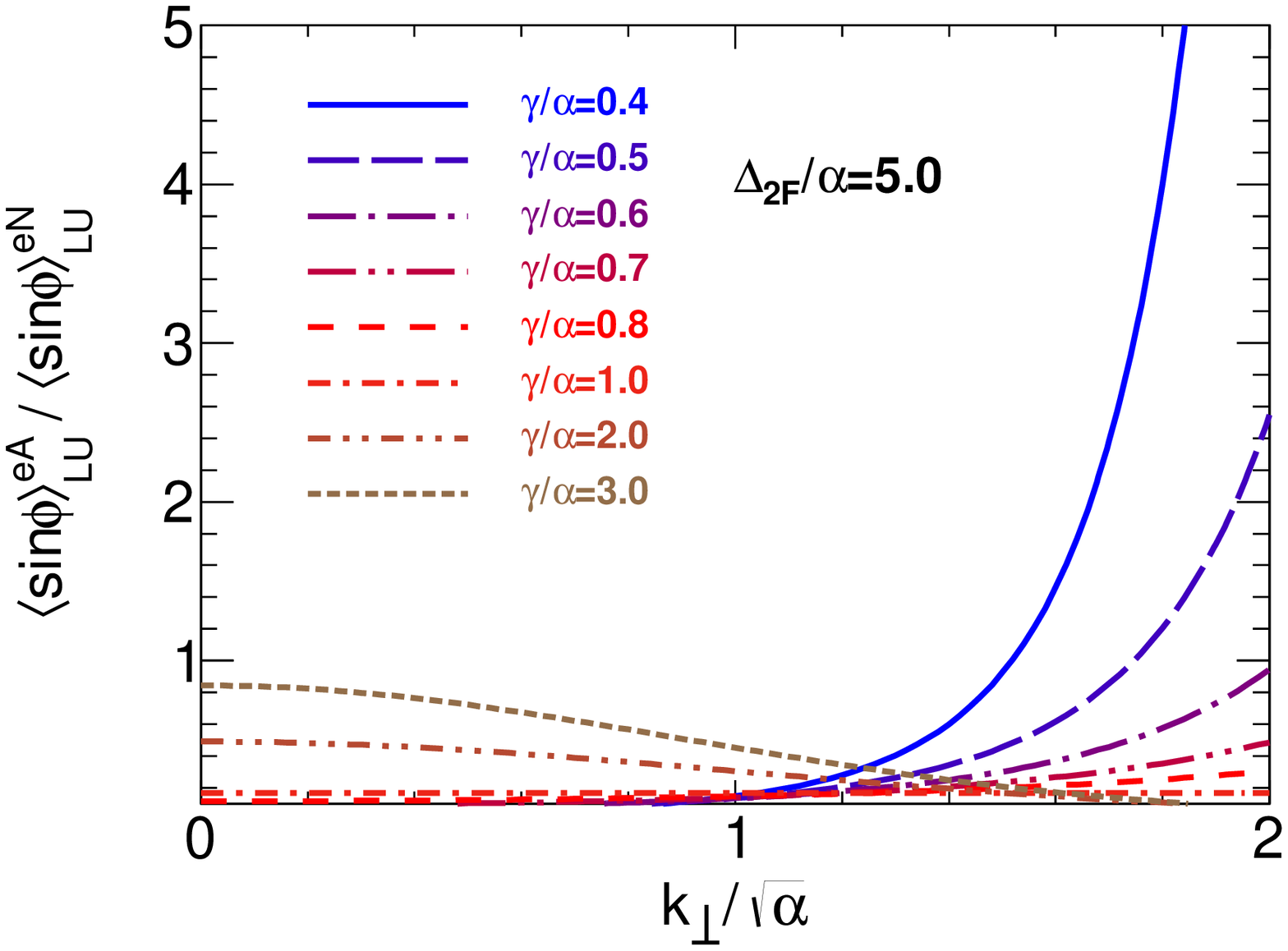,width=0.55\textwidth,clip=}
 \caption{Ratio of $\langle\sin\phi\rangle_{LU}^{eA}/\langle\sin\phi\rangle_{LU}^{eN}$ as a function of $k_\perp$ for different $\gamma$ and $\Delta_{2F}$.}
\end{figure}

\section{Summary}

We present a systematic calculation of the hadronic tensor and azimuthal asymmetries 
in the semi-inclusive deep-inelastic scattering $e^-+N\to e^-+q+X$ with polarized beam and/or polarized target 
based on the collinear expansion formalism in LO pQCD and up to twist-3 contributions.  
The results depend on a number of new TMD parton correlation functions.
We showed that measurements of the corresponding azimuthal asymmetries and their $k_\perp$-dependence 
can provide much information on these TMD correlation functions which in turn can shed light on the properties 
of multiple gluon interaction in hadronic processes. 
We presented the results also for reactions with nucleus target $e^-+A\to e^-+q+X$ 
and discuss the nuclear dependence. 
We show that the relationship between these TMD correlation functions inside large nuclei and that of a nucleon 
under two-gluon correlation approximation. 
One can study the nuclear dependence of the different azimuthal asymmetries which
are determined by the corresponding parton distribution and correlation functions. 
With the Gaussian ansatz for the TMD parton correlation functions inside the nucleon, 
we also illustrate numerically that the asymmetries are suppressed 
in the corresponding SIDIS with nuclear target.

Experimental studies of the azimuthal asymmetries have been carried out in both unpolarized and polarized 
semi-inclusive deep inelastic scattering with nucleon 
target~\cite{Aubert:1983cz,Arneodo:1986cf,Adams:1993hs,Breitweg:2000qh,Chekanov:2002sz,
Airapetian:2001eg,Airapetian:2002mf,Airapetian:2004tw,Alexakhin:2005iw,Webb:2005cd,Osipenko:2008aa,Alekseev:2010dm}. 
More results are expected from CLAS at JLab and COMPASS at CERN.
The available data seem to be consistent with the Gaussian ansatz for the 
transverse momentum dependence of the TMD matrix elements in the unpolarized case~\cite{Schweitzer:2010tt}.
However these data are still not adequate enough to provide any precise  constraints on the form of 
the higher twist matrix elements.  
Our calculations of the azimuthal asymmetries are
most valid in the small transverse momentum region where NLO pQCD corrections are not dominant. 
The high twist effects are also most accessible in intermediate region of $Q^{2}$.  
One expects that future experiments such as those at the proposed Electron Ion 
Collider (EIC)~\cite{Boer:2011fh} will be better equipped to study these high twist effects in detail.

This work was supported in part by the National Natural Science Foundation of China 
under grant No. 11035003, No. 11105137,  and  No. 11221504.
Office of Energy Research, Office of High Energy and Nuclear Physics, 
Division of Nuclear Physics, of the U.S. Department of Energy under Contract No.
DE-AC02-05CH11231, and by CCNU-QLPL Innovation Fund (QLPL2011P01) .


\begin{thebibliography}{}
\bibitem{ff1} The notations that we use here are slightly different from those in [9]. 
The $W$'s defined here are equivalent to the $F$'s in [9] multiplied by the corresponding sine or cosine.
\bibitem{ff2} We see also that the first and the second terms in $W_{UT}$, the $W_{LU}$, the $W_{UL}$, 
and the third term in $W_{LT}$ have different signs from the corresponding terms of the results presented in [9].  
This is caused by the difference in the definition of $\phi$ and $\phi_s$ in the two papers. 
In [9], when defining $\sin\phi$ and $\sin\phi_s$, $z$-direction is taken as 
the direction of the momentum of the virtual photon. 
However, in defining $x=k^+/p^+$, the 4-momentum $p$ of the incident nucleon is taken in a way that 
the $+$ component in the lightcone coordinate dominates which implies 
that the $z$-direction is taken as the direction of motion of the nucleon. 
In this paper, we unify the notation by taking $z$-direction as the direction of motion of the incident nucleon 
both in defining $x$ and $\sin\phi$ thus we define $\sin\phi$ and $\sin\phi_s$ as mentioned after Eq.(69) 
and they have a sign difference with those in [9].
\end{thebibliography}
\end{document}